\documentclass{mn2e}
\usepackage{epsfig,natbib2,natbibmnfix}
\usepackage{amssymb}
\usepackage{amssymb}
\usepackage{multirow}
\usepackage{multicol}
\usepackage[varg]{txfonts}
\usepackage{color}

\newcommand{\cref}[1]{Chapter~\ref{#1}}

\begin{document}
\title[A photospheric metal line profile analysis of hot DA white dwarfs with
  circumstellar material]{A photospheric metal line profile analysis of hot DA white dwarfs with
  circumstellar material}

\author[N.J.~Dickinson et al.] 
{
\parbox{5in}{N.J. Dickinson$^{1}$, M.A. Barstow$^{1}$ and B.Y. Welsh$^2$}
\vspace{0.1in} 
\\ $^{1}$Department of Physics \& Astronomy, University of Leicester,
 Leicester LE1 7RH. UK.
\\ $^2$ Space Sciences Laboratory, University of California, Berkeley, California, USA
}

\maketitle

\begin{abstract}
Some hot DA white dwarfs have circumstellar high ion absorption
features in their spectra, in addition to those originating in the
photosphere. In many cases, the line profiles of these absorbing
components are unresolved. Given the importance of the atmospheric
composition of white dwarfs to studies of stellar evolution,
extra-solar planetary systems and the interstellar medium, we examine
the effect of including circumstellar line profiles in the abundance
estimates of photospheric metals in six DA stars. The photospheric
C and Si abundances are reduced in five cases where the
circumstellar contamination is strong, though the relative weakness of
the circumstellar Si\,{\sc iv} absorption introduces minimal
contamination, resulting in a small change in abundance. The inability
of previous, approximate models to reproduce the photospheric line
profiles here demonstrates the need for a technique that accounts for the physical line profiles of both the circumstellar and photospheric lines when modelling these blended absorption features.
\end{abstract}


\footnotetext[1]{E-mail: njd15@le.ac.uk}

\section{Introduction}
\label{intro}

White dwarfs are the evolutionary end products of most stars. As such,
a detailed knowledge of these objects is crucial to comprehend the end states of
stellar evolution and planetary systems. Accurate \textit{T}$\rm _{eff}$ and
log \textit{g} measurements for white dwarfs are necessary for a proper
understanding of these bodies. Using evolutionary models
(e.g. \citealt{FontaineBrassardBergeron01}), these stellar parameters
can be used to derive a mass, which can then be used in studies of white dwarf
mass distributions
(e.g. \citealt{BergeronSafferLiebert92,LiebertBergeronHolberg05}),
initial-final mass relations (e.g. \citealt{Casewelletal09,Dobbieetal09}) or
luminosity functions (e.g. \citealt{LiebertBergeronHolberg05}). Given that
white dwarfs represent some of the oldest stellar objects, their ages can be used to date stellar populations, or the Galactic disc (e.g. \citealt{FontaineBrassardBergeron01}). This makes reliable measurements of stellar parameters such as \textit{T}$\rm _{eff}$ and log \textit{g} (found by fitting the Balmer/Lyman series in DA white dwarfs; e.g. \citealt{Holbergetal85,BergeronSafferLiebert92}), and thus a robust understanding of the white dwarf atmospheres in which the absorption series arise, critical.

Over the previous few decades of white dwarf research, evidence for the
presence of metals in the photospheres of hot DA stars has abounded
(e.g. \citealt{Barstowetal93,Barstowetal03,Marshetal97}), where radiative
levitation is sufficient to counter the downward diffusion of the heavy ions
(e.g. \citealt{Chayeretal94}; \citealt*{Chayeretal95a}; \citealt{Chayeretal95b}). Proper inclusion of
these metals significantly affects the predicted Balmer/Lyman line profiles,
influencing the measured white dwarf \textit{T}$\rm _{eff}$ and log
\textit{g}; the measured \textit{T}$\rm _{eff}$ for DAs hotter than 55\,000\,K
(where radiative levitation is the dominant process governing metallic
composition, and metal line blanketing affects \textit{T}$\rm _{eff}$
measurements; e.g. \citealt{DreizlerWerner93}) using a metallic, non-local
thermodynamic equilibrium (NLTE) analysis reduces the measured \textit{T}$\rm
_{eff}$ by 4\,000 to 7\,000\,K when compared to a pure hydrogen, local
thermodynamic equilibrium (LTE) analysis
(\citealt{BarstowHubenyHolberg98}). Furthermore, for objects with
\textit{T}$\rm _{eff} >$ 50\,000\,K, the \textit{T}$\rm _{eff}$ values measured
using Balmer and Lyman lines are inconsistent for a given object by up
to 10\,000\,K
(\citealt{Barstowetal01,Barstowetal03BalLy}). This is also observed at DAO
stars (\citealt{Goodetal04}), with more severity (with \textit{T}$\rm _{eff}$ values above
55\,000\,K differing by up to 60\,000\,K in some cases). Thus, properly accounting for the metals in hot
white dwarfs is crucial to a proper understanding of the \textit{T}$\rm
_{eff}$ and log \textit{g} of a given star.

However, the inclusion of metals in hot white dwarf models at the
correct abundance is not trivial. Using homogeneous models of white dwarf photospheres, \cite{Barstowetal03}
found some significant deviations between the measured abundances and those
predicted by radiative levitation for the stars in their sample, though the
predicted abundance-\textit{T}$_{\rm eff}$/log \textit{g} patterns are
reproduced in that hotter stars have higher metal abundances. Indeed, some stars with similar \textit{T}$_{\rm eff}$ and log
\textit{g} values have
somewhat different metal abundances. In addition to homogeneously distributed
models, self-consistent stratified models, where the metal abundances present
are those resulting from diffusive equilibrium, have been used to model hot DA
stars (\citealt{Dreizler99,DreizlerWolff99,SchuhDreizlerWolff02}). Comparisons
of the abundances measured using the stratified  models of
\cite{SchuhDreizlerWolff02} to the homogeneous models of \cite{Barstowetal03}
show that, although abundance measurements are roughly consistent with
each other across the sample (when systematic errors are accounted for),
measurements do not match up on an object by object basis
(\citealt*{SchuhBarstowDreizler05}). Models with highly abundant metals present in only the upper atmospheric regions have been used to explain the profiles
of the N\,{\sc v} (\citealt{Holbergetal95,Holbergetal99ph,Holbergetal00,Barstowetal03}) and O\,{\sc vi} \citep{Chayeretal06} absorption lines in some stars, while other
studies find homogeneous metal distributions with abundances in keeping with
those of DAs with higher \textit{T}$_{\rm eff}$ represent the observations well
\citep{Chayeretal05,Dickinsonetal12nv}. Models with stratified Fe have also
been used to explain observations of WD\,0501+527
(G191-B2B; \citealt*{BarstowHubenyHolberg99}; \citealt{Dreizler99,DreizlerWolff99}). Furthermore, absorption from elements such as Ge, Sn,
Pb \citep{VennesChayerDupuis05} and Ar (\citealt*{WernerRauchKruk07}) has been
detected in DA stars.

At cooler white dwarfs, where metallic ions should sink out of the stellar
atmospheres on short timescales (e.g. \citealt{KoesterWilken06}), photospheric
metals, where observed, are attributed to the accretion of tidally
disrupted minor planets or asteroids
(e.g. \citealt{Zuckermanetal07,Zuckermanetal11,Dufouretal10,Kleinetal11}),
since white dwarf kinematics and abundances do not favour ISM accretion (\citealt{Aannestadetal90,ZuckermanReid98,Zuckermanetal03,Farihietal10}) 
and the C, Si and Al abundances observed in the sample studied by \cite{Dupuisetal10} require measurable accretion in
addition to the recently calculated radiative levitation effects (\citealt{ChayerDupuis10}). Circumstellar discs with dust
and gas components have been seen around some hot DAs (\citealt{Gaensickeetal06,Gaensickeetal07,Gaensickeetal08,Brinkworthetal09,Melisetal11,Debesetal12}), with the gaseous components being due to the sublimation of
dust grains by the intense radiation from the hot stars. The anomalously high metal abundances
seen in some hot DA stars (e.g. WD\,2111+498/GD\,394;
\citealt{Holbergetal97,Chayeretal00,Dupuisetal00,Vennesetal06}) may be similarly
linked to the accretion of circumstellar material, most likely in the
form of a gas disc, due to the sublimation of dust expected near white
dwarfs with higher \textit{T}$_{\rm eff}$ values. However, searches for such
material around hot DAs with anomalous metal abundances have not yet
yielded any definitive detections of either gas disc emission or
infrared excesses
(e.g. \citealt{Burleighetal10,Burleighetal11}, \textit{in preparation}). A
proper understanding of the hot DAs with anomalous metal abundances is crucial to our understanding of which hot stars may be accreting, and will therefore impact significantly on our understanding of the fate of planetary systems at the evolutionary end point of their host stars.
 
Further complications arise in our understanding of metal absorption in these
objects, when non-photospheric high ion absorption features are present in DA
spectra. Such absorption features were first detected nearly 20 years ago in the
\textit{IUE} spectrum of WD\,1620$-$391 (CD\,$-$38$\rm ^o$10980), at velocities far from the
the photospheric and interstellar medium (ISM) lines, leading to the
interpretation that they were
`circumstellar' (\citealt{Holbergetal95}). Later studies have found such
circumstellar features in other white dwarf spectra
(\citealt{Holbergetal97,HBS98,Bannister03,Lallementetal11}). Originally
thought to be related to stellar mass loss
(\citealt{HBS98,HBS99,Bannister03}), vaporised planetesimals
(\citealt{Lallementetal11}) or the ionisation of either the ISM or
ancient, diffuse remnants of planetary nebulae (PNe) (\citealt{Bannister03,Dickinsonetal12circ}) have
been proposed as alternative origins for the observed features, since mass loss is no longer thought to occur
at these hot DAs (\citealt{Dickinsonetal12circ,Unglaub07,Unglaub08}). Indeed,
the ionisation of the ISM inside hot white dwarf Str\"{o}mgren spheres may
account for some of the observed ionisation structure of the LISM
(e.g. \citealt{Welshetal10}, Welsh et al. \textit{in preparation}). Furthermore, non-photospheric
Si\,{\sc iv} absorption has been identified in \textit{Hubble Space
  Telescope} (\textit{HST}) Cosmic Origins Spectrograph (COS) observations of WD\,0843+516
(PG\,0843+516) and SDSS\,1228+1040 (given the strong Si\,{\sc iv} detections and lack of C\,{\sc iv} at these objects, these features are attributed
to absorption in the circumstellar discs about the stars, and are not
thought to be from the same type of absorber as those seen at the hotter DAs; \citealt{Gaensickeetal12}). 

In some cases, these
circumstellar features have velocities close to those of the
photospheric absorption lines, leading to blended absorption line
profiles. Some studies (e.g. \citealt{Barstowetal03}) that estimated the metal abundances of stars where such
blended absorption lines are present neglected the circumstellar
components of the absorption features, since the potential for the
circumstellar components to act as contaminants to photospheric
abundance measurements was not then know. However,
\cite{Dickinsonetal12circ} found that the circumstellar
components to the 1548 \AA\ and 1550 \AA\ C\,{\sc iv} features in the spectrum
of WD\,0501+527 account for a significant proportion of the absorption line
profiles, having equivalent widths of 128.77 m\AA\ and 106.47 m\AA\
respectively, while the total equivalent widths of the features are 160.01
m\AA\ and 141.93 m\AA\, demonstrating the sizeable effect the
circumstellar absorption may have on photospheric abundance estimates.

The recent study of DA circumstellar absorption by \cite{Dickinsonetal12circ}
has allowed, for the first time, full characterisation of the circumstellar
absorbing components (using the technique described in studies such as
\citealt{WelshLallement05,WelshLallement10}). By taking proper account of
these non-photospheric absorption profiles in blended high ion features, more
robust photospheric metal abundances can be derived. Given the great importance and wide
ranging applications a thorough understanding of white dwarf metal
content has, we present here an analysis of the photospheric components
of such blended absorption features, with the aim of better
understanding how the stellar absorption features contribute to the blended
absorption line profiles, and to see how the inclusion of the previously
characterised circumstellar components affects the derived DA metal
abundances.

\section{Observations and Method}
\label{method}

\begin{table*}
\caption{The stellar parameters of the white dwarfs with blended high ion
  absorption features, and the observation information for the data used.}
\begin{tabular}{l c c c c c c c}
\hline
WD             & Alt.name        & \textit{T}$_{\rm eff}$ &
log\,\textit{g} & Data source [Mode]    & Resolving Power
($\lambda/\Delta\lambda$, FWHM)\\
\hline   
0232+035$^*$   & Feige\,24       & 61\,000$\pm$1\,100    & 7.50$\pm$0.06  & \textit{STIS} [E140M] & 40\,000\\
0501+527       & G191-B2B        & 53\,500$\pm$900       & 7.53$\pm$0.09 & \textit{STIS} [E140H] & 110\,000\\
0556$-$375     & REJ\,0558$-$373 & 60\,000$\pm$2\,200    & 7.70$\pm$0.09 & \textit{STIS} [E140M] & 40\,000\\
0939+262       & Ton\,021        & 69\,700$\pm$530       & 7.47$\pm$0.05 & \textit{STIS} [E140M] & 40\,000\\ 
1611$-$084     & REJ\,1614$-$085 & 38\,800$\pm$480       & 7.92$\pm$0.07 & \textit{GHRS} [G160M] & 22\,000\\
2218+706       &                 & 58\,600$\pm$3\,600    & 7.05$\pm$0.12 & \textit{STIS} [E140M] & 40\,000\\
\hline                
\end{tabular}
\label{table:tefflogg}
 \\\footnotesize{$^*$the blending of components occurs only at the
   0.24 binary phase. Note: though the \textit{T}$_{\rm eff}$ and
log\,\textit{g} values here are taken from \cite{Barstowetal03}, they
have been adjusted to reflect the significance of the stated error.}
\end{table*}

Of the 23 stars in the sample analysed by \cite{Bannister03} and
\cite{Dickinsonetal12circ}, eight yield unambiguous circumstellar
detections. At only two of these objects (WD\,0455$-$282 and WD\,1738+665) are the circumstellar
features completely resolved from their photospheric counterparts, while at WD\,0232+035
the components are resolved at one of the binary phases (0.74). Table
\ref{table:tefflogg} lists the objects at which the circumstellar and
photospheric components are blended, and gives the stellar parameters of the
objects (from \citealt{Barstowetal03}). We used data from the
\cite{Barstowetal03} study\footnote[1]{available from the MAST
  archive (http://archive.stsci.edu)}, with observation information
for each DA in table \ref{table:tefflogg}.

\begin{table*}
\caption[]{The circumstellar C\,{\sc iv} and Si\,{\sc iv} laboratory
  wavelengths ($\lambda_{\rm lab}$), oscillator strengths  ($f$),
  measured line parameters ($v\rm _{circ}$, \textit{b}, \textit{N}),
  model input parameter values ($E_l$, $\sigma_{l}$, $l_{\rm d}$) and $v \rm _{phot}$ values for the white dwarfs studied here.}
\scriptsize{
\begin{tabular}{l c c c c c c c c c}
\hline
\multicolumn{10}{c}{C\,{\sc iv}}\\
\hline
WD                               & $\lambda_{\rm lab}$ (\AA)  & $f^\dagger$       &  $v\rm _{circ}$  (km s$^{-1}$) &\textit{b} (km s$^{-1}$)  & \textit{N} (x10$\rm ^{12} cm^{-2}$)    &   $E_l$ (x10$^{-3}$ keV) & $\sigma_{l}$ (x10$^{-7}$ keV) & $l_{\rm d}$ (x10$^{-7}$) & $v\rm _{phot}$ (km s$^{-1}$)\\
\hline   
0232+035$^{\rm a}$     & 1548.187                          & 0.19     &  7.3                                   & 6.2              & 27.6                                   &   8.008                  &  1.171                            & 5.746                    &  125.7\\
                                    & 1550.777                          & 0.0952 & 7.1                                    & 6.5              & 29.0                                   &   7.995                  &  1.226                           & 3.027                    & 128.9\\
0232+035$^{\rm b}$     & 1548.187                          & 0.19     &  7.9                                   & 7.1              & 29.4                                   &   8.008                 & 1.341                            & 6.132                    & 29.7\\
                                   & 1550.777                           & 0.0952 & 7.1                                   & 5.8              & 26.8                                    &   7.995                  &  1.094                           & 2.803                   & 28.7\\
0501+527                   & 1548.187                           & 0.19    &  8.0                                   & 5.9              & 107.0                                  &   8.008                 &  1.114                            & 22.221                    & 26.7\\
                                    & 1550.777                          & 0.0952 & 9.8                                   & 5.4              & 101.0                                   &    7.995                &  1.018                           & 10.534                  & 27.0\\
0556$-$375                & 1548.187                          & 0.19     &  9.1                                  & 11.0               & 39.2                                  &   8.008                  &  2.078                            & 8.174                    & 32.3\\
                                    & 1550.777                          & 0.0952 & 11.0                                 & 1.5              & 54.4                                    &    7.995                 &  2.168                            & 5.131                  & 31.5\\
0939+262                   & 1548.187                           & 0.19     &  10.0                                & 8.3              & 7.53                                    &   8.008                 & 1.568                             & 1.571                    & 35.2\\
                                    & 1550.777                          & 0.0952 & 7.9                                   & 8.3              & 8.57                                    &    7.995                  & 1.565                            & 0.895                  & 34.2\\
1611$-$084                & 1548.187                          & 0.19     &  $-$17.7                          & 3.7              & 6.57                                    &  8.008                    & 0.699                            & 1.371                     & $-$37.4\\
                                    & 1550.777                          & 0.0952 & $-$58.2                           & 3.7              & 2.3                                      &   7.997                   & 0.698                           & 2.399                  & $-$34.6\\
2218+706                   & 1548.187                          & 0.19      &  $-$17.4                          & 5.0              & 118                                   &  8.008                   & 0.945                            & 24.702                      & $-$40.5\\
                                    & 1550.777                         & 0.0952  & $-$18.1                            & 8.5              & 120                                  &    7.995                  & 1.603                             & 12.498                   & $-$42.5\\
\hline
\multicolumn{10}{c}{Si\,{\sc iv}}\\
\hline   
0939+262                & 1393.755                            & 0.513    & 6.2                                       & 9.6              & 1.34                                  &    8.955                     & 2.401                             & 0.753                  & 36.4\\
                                 & 1402.770                           & 0.225    & 18.4                                & 13.7           & 2.28                                  &    8.838                  & 2.856                            & 0.639                 & 38.0\\
1611$-$084             & 1393.755                           & 0.513    & $-$22.8                            & 2.4             & 1.30                                  &    8.896                  & 2.560                             & 0.731                  & $-$47.9\\
2218+706                & 1393.755                           & 0.513    & $-$18.5                            & 9.2              & 7.34                                    &    8.896                  & 1.930                             & 4.133                  & $-$40.7\\
                                & 1402.770                            & 0.225    & $-$17.2                           & 12.9             & 8.36                                   &    8.839                  & 2.856                             & 2.415                  & $-$40.4\\
\hline 
\end{tabular}
}
\label{table:fitpars}
\\\footnotesize{$^\dagger$from the NIST Atomic Spectra Database,
  $^a$resolved spectrum (0.74 binary phase), $^b$blended spectrum
  (0.24 binary phase). Note: no errors presented, since $E_l$, $\sigma_{l}$ and $l_{\rm d}$ were allowed to vary freely, not within a hard range.}
\end{table*}

The method used to fit the spectra is a variation of the method used in
previous work (e.g. \citealt{Barstowetal03, Dickinsonetal12nv}), so a summary
is presented here with an explanation of how we accounted for the circumstellar components. {\sc xspec}  (\citealt{Arnaud96}) was utilised to perform the
spectrum fitting. The \textit{T}$_{\rm eff}$ and log
\textit{g} values from table \ref{table:tefflogg} were used in the fits, with the parameters
allowed to vary within the stated error range. The circumstellar
components of the absorption lines were accounted for using the {\sc xspec}
Gaussian absorption line model `{\sc gabs}'. The circumstellar line properties (circumstellar line
velocity, $v\rm _{circ}$; \textit{b}
value; column density, \textit{N}) measured by
\cite{Dickinsonetal12circ} were converted to the quantities used as
input parameters to {\sc gabs} (line energy in keV, $E_l$; Gaussian
sigma in keV, $\sigma_{l}$; line depth, $l_{\rm d}$)  using equations
\ref{eq:El}-\ref{eq:ld}, and used as starting values for the Gaussian
components (table 
\ref{table:fitpars}).

\begin{equation}
E_l = \frac{hc}{\lambda_{\rm lab}\left ( 1+\frac{v_{\rm circ}}{c} \right )}
\label{eq:El}
\end{equation}
\begin{equation}
\sigma_l = \frac{hc}{\sqrt{2}\lambda _{\rm lab}}\left ( \frac{1}{1+\frac{v_{\rm circ}}{c}}-\frac{1}{1+\frac{v_{\rm circ}+b}{c}} \right )
\label{eq:sig}
\end{equation}
\begin{equation}
l_{\rm d} = \sqrt{2\pi}\sigma_l\tau
\label{eq:ld}
\end{equation}
where $\lambda_{\rm lab}$ is the laboratory wavelength of the
absorption line, $c$ is the speed of light, $h$ is Planck's constant
and $\tau$ is the optical depth at the line centre, calculated using
the formalism of \cite{Spitzer78} (equation \ref{eq:optdepth}).
 
\begin{equation}
\tau =  \frac{N\sqrt{\pi}e^{2}f}{m_{e}c}\frac{\lambda}{b}
\label{eq:optdepth}
\end{equation}
where $e$ and $m_e$ are the charge and mass of an electron and $f$ is the oscillator strength.

The parameters of the Gaussian absorption lines were allowed
to vary freely, since the fitting method here differs from that used 
by \cite{Dickinsonetal12circ}, and thus the errors associated with
$v\rm _{circ}$, \textit{b}  and \textit{N}  were not considered in the calculation
of $E_l$, $\sigma_{l}$ and $l_{\rm d}$, and are therefore not presented
in table \ref{table:fitpars}. The two spectra of WD\,0232+035 were fitted separately, not summed in the white dwarf rest frame as
in previous work (e.g. \citealt{Barstowetal03,VennesLanz01}), to allow a consistency check
of the abundances derived with and without the circumstellar component
contaminating the C\,{\sc iv} line profiles.

The photospheric absorption components were modelled using the NLTE stellar
atmosphere code {\sc tlusty} (\citealt{HubenyLanz95}). The wavelength regions used to fit the
spectra are the same as those used by  \cite{Barstowetal03}, namely
1545 to 1555\,\AA\ for the C\,{\sc iv} doublet and 1390 to 1405\,\AA\ for the Si\,{\sc iv} doublet. The absorption lines from any element other
than that considered in each spectral region were excluded to avoid the coupling
of the abundances of multiple elements. These `background' elements were
included at the abundances stated in \cite{Barstowetal03}. The line
centres of the photospheric components were initially placed at the photospheric velocities ($v\rm _{phot}$,  table
\ref{table:fitpars}) found
by \cite{Dickinsonetal12circ}, and allowed to fit freely. 1$\sigma$
errors were computed for all abundance measurements.

\section{Results}

\begin{figure}
\includegraphics[angle=90, width=0.47\textwidth]{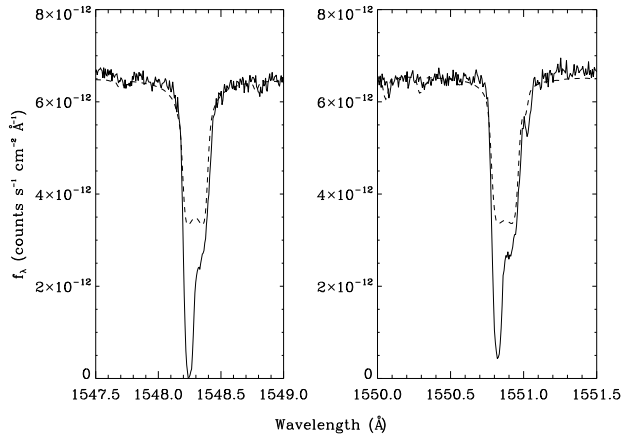}
 \caption{The  C {\sc iv} doublet of WD\,0501+527, fit with a model spectrum with the
   C abundance (C/H = 4.00x10$\rm^{-7}$) derived from the C\,{\sc iv} doublet by
   \cite{Barstowetal03}. The observed data is shown with a solid line,
   while the model is plotted with a dashed line.}
  \label{fig:MAB}
\end{figure}

\begin{figure}
\includegraphics[angle=90, width=0.47\textwidth]{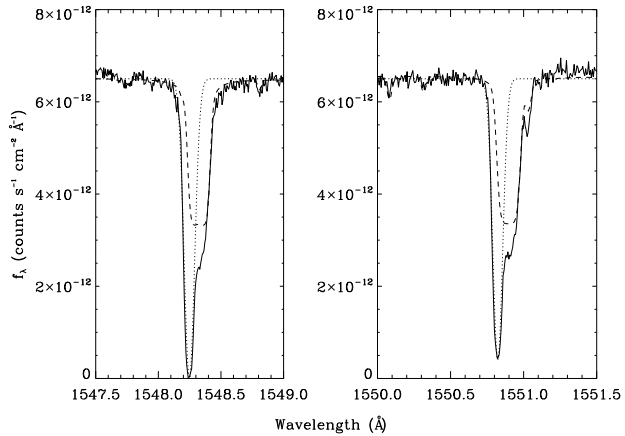}
 \caption{The best fitting model of the C {\sc iv} doublet of WD\,0501+527, with a photospheric C abundance of
 1.4x10$\rm^{-7}$ relative to hydrogen (dashed line). The dotted line represents the 
 circumstellar component and the solid line
 is the observed spectrum.}
  \label{fig:BFfit}
\end{figure}

\begin{figure*}
\includegraphics[angle=90, height=10cm]{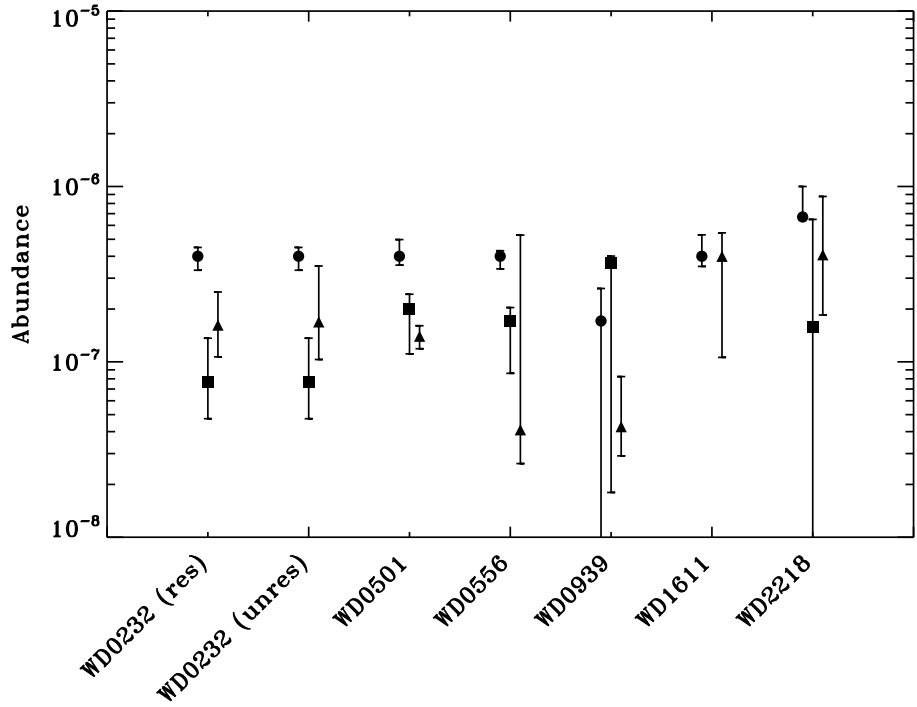}
  \caption{The C abundances measured in this study using the C\,{\sc iv}
    doublet (triangles), compared to the abundances derived using the  C\,{\sc iii} (squares) and C\,{\sc iv} (circles) absorption features by \cite{Barstowetal03}.}
  \label{fig:civ_teff}
\end{figure*}

\begin{figure*}
\includegraphics[angle=90, height=10cm]{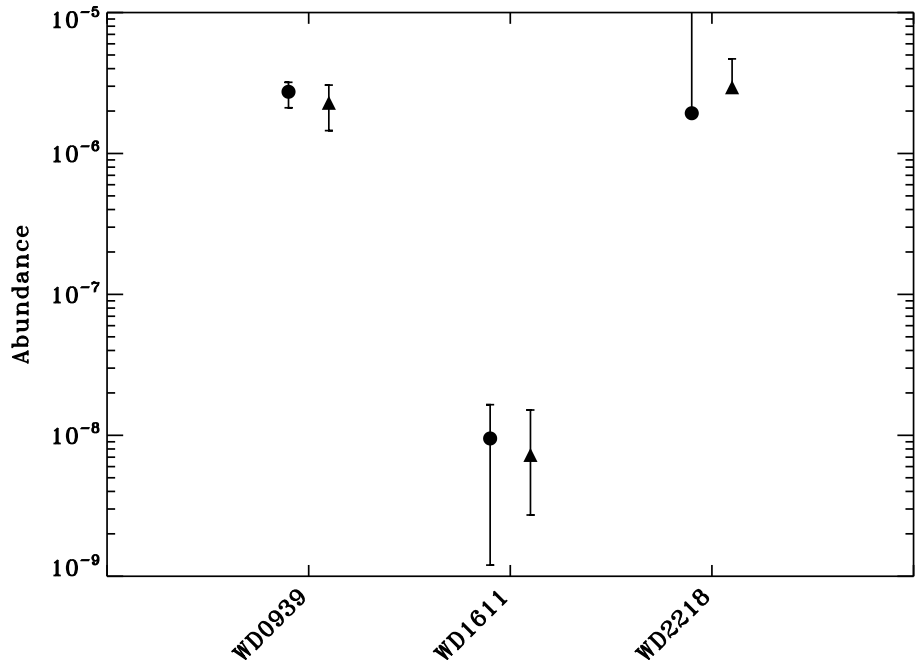}
  \caption{The Si abundances measured in this study (triangles), compared to the abundances derived by \cite{Barstowetal03} (circles).}
  \label{fig:siiv_teff}
\end{figure*}

Table \ref{table:newabs} details the new abundances found in this
study, with the abundances found by \cite{Barstowetal03} presented for
comparison.  The improvement seen progressing from a photospheric only
model (figure \ref{fig:MAB}) to a model that contains photospheric and
circumstellar components in the C\,{\sc iv} doublet of WD\,0501+527 is illustrated in figure \ref{fig:BFfit}. Figure \ref{fig:civ_teff} shows a
comparison of the C abundances derived here to those found by
\cite{Barstowetal03}, and figure \ref{fig:siiv_teff} illustrates how
the Si abundances measured here compare to previous estimates. The C
abundances derived here using the C\,{\sc iv} doublet have been
revised down when compared to those from \cite{Barstowetal03} for all
but two objects (WD\,1611$-$084 and WD\,2218+706). The C abundances obtained from the different
spectra of WD\,0232+035 agree well, demonstrating that even in
co-added spectra circumstellar contamination can be significant. The
Si\,{\sc iv} absorption lines do not suffer from such strong
circumstellar contamination, leading to less
revision of the Si abundances. 

\begin{table*}
\caption{The abundances measured in this study, with 3$\sigma$ uncertainties. The C and Si abundances found by \cite{Barstowetal03} are also stated for comparison.}
\begin{tabular}{l c c c c c c c c c}
\hline
WD                 & C\,{\sc iv}/H$^{\dagger}$  & $+$1$\sigma$        & $-$1$\sigma$       & C\,{\sc iii}/H$^{\ddagger}$  & $+$3$\sigma$        & $-$3$\sigma$       & C\,{\sc iv}/H$^{\ddagger}$       & $+$3$\sigma$       & $-$3$\sigma$\\
\hline   
0232+035$^{\rm a}$  & 1.62$\times10^{-7}$       & 8.80$\times10^{-8}$ & 5.53$\times10^{-8}$ & 7.64$\times10^{-8}$          & 6.00$\times10^{-8}$ & 2.90$\times10^{-8}$ & 4.00$\times10^{-7}$ & 6.70$\times10^{-8}$ & 4.90$\times10^{-8}$\\
0232+035$^{\rm b}$  & 1.69$\times10^{-7}$       & 1.83$\times10^{-7}$ & 6.59$\times10^{-8}$ & 7.64$\times10^{-8}$          & 6.00$\times10^{-8}$ & 2.90$\times10^{-8}$ & 4.00$\times10^{-7}$ & 6.70$\times10^{-8}$ & 4.90$\times10^{-8}$\\
0501+527           & 1.40$\times10^{-7}$       & 2.07$\times10^{-8}$ & 2.12$\times10^{-8}$ & 1.99$\times10^{-7}$          & 4.40$\times10^{-8}$ & 8.80$\times10^{-8}$ & 4.00$\times10^{-7}$ & 4.40$\times10^{-8}$ & 9.80$\times10^{-8}$\\
0556$-$375         & 4.10$\times10^{-8}$       & 4.87$\times10^{-7}$ & 1.47$\times10^{-8}$ & 1.70$\times10^{-7}$          & 8.40$\times10^{-8}$ & 3.40$\times10^{-7}$ & 4.00$\times10^{-7}$ & 6.10$\times10^{-8}$ & 3.00$\times10^{-8}$\\
0939+262           & 4.27$\times10^{-8}$       & 3.97$\times10^{-8}$ & 1.36$\times10^{-8}$ & 3.68$\times10^{-7}$          & 3.30$\times10^{-7}$ & 3.50$\times10^{-7}$ & 1.71$\times10^{-7}$ & 1.70$\times10^{-8}$ & 9.10$\times10^{-8}$\\
1611$-$084         & 4.00$\times10^{-7}$       & 1.44$\times10^{-7}$ & 2.94$\times10^{-7}$ &                             &                     &                    & 4.00$\times10^{-7}$ & 5.00$\times10^{-8}$ & 1.30$\times10^{-7}$\\
2218+706           & 4.08$\times10^{-7}$       & 2.23$\times10^{-7}$ & 4.96$\times10^{-7}$ & 1.59$\times10^{-7}$          & 4.90$\times10^{-7}$ & 1.58$\times10^{-7}$ & 6.70$\times10^{-7}$ & 2.80$\times10^{-6}$ & 3.30$\times10^{-7}$\\
\hline
WD                 & Si\,{\sc iv}/H$^{\dagger}$ &  $+$1$\sigma$       & $-1\sigma$         & Si\,{\sc iv}/H$^{\ddagger}$      & $+$3$\sigma$       & $-$3$\sigma$\\
\hline   
0939+262           & 2.96$\times10^{-6}$       & 7.67$\times10^{-7}$ & 8.37$\times10^{-7}$ & 2.74$\times10^{-6}$ & 6.30$\times10^{-7}$ & 4.60$\times10^{-7}$\\
1611$-$084         & 7.26$\times10^{-9}$       & 7.87$\times10^{-9}$ & 4.54$\times10^{-9}$ & 9.50$\times10^{-9}$ & 7.00$\times10^{-9}$ & 8.30$\times10^{-9}$\\
2218+706           & 5.08$\times10^{-7}$       & 1.74$\times10^{-6}$ & 2.33$\times10^{-7}$ & 1.93$\times10^{-6}$ & 1.10$\times10^{-5}$ & 1.10$\times10^{-6}$\\
\hline                 
\end{tabular}
\label{table:newabs}
\\\footnotesize{$^{\dagger}$values from this study, $^{\ddagger}$values from
\cite{Barstowetal03}, $^a$resolved spectrum (0.74 binary phase),
$^b$blended spectrum (0.24 binary phase). Note: while 1$\sigma$ errors were computed here,
  \cite{Barstowetal03} computed 3$\sigma$ errors to allow the
  estimation of upper abundance limits where no unambiguous absorption
lines were seen in their study, and so they are presented here.}
\end{table*}

\section{Discussion}
\label{discussion}

The downward revision of the C abundances of the stars examined
here shows the significant effect circumstellar contamination can have
on photospheric metal abundance measurements. Indeed, the C
abundance for WD\,0501+527 derived here using the C\,{\sc iv} doublet
is in better agreement with that measured by \cite{Barstowetal03}
using the C\,{\sc iii} lines than that found using the C\,{\sc iv}
doublet. Furthermore, the abundance derived here is
also consistent with the C abundance derived by
\cite{VennesLanz01} for WD\,0501+527, and with that measured using the
C\,{\sc iii} multiplet in the star's \textit{FUSE} spectrum (Barstow
et al., \textit{in preparation}). The C and Si abundances of
WD\,1611$-$084  have been subject to little change, with the C
abundance consistent with that derived using \textit{FUSE} data; the Si abundance measured
from the \textit{FUSE} spectrum (Barstow et al., \textit{in
  preparation}) is an order of magnitude larger than that found both here
and by \cite{Barstowetal03}, suggesting that circumstellar absorption
may also be present in the \textit{FUSE} data. Interestingly, the photospheric
abundances of this star are in excess of those of most DAs with similar
\textit{T}$\rm _{eff}$ values, and when coupled with the difficulties in both modelling
the distribution of nitrogen in this star
(\citealt{Holbergetal99ph,Chayeretal05,Dickinsonetal12nv}) and in ascertaining
the origin of its circumstellar material (\citealt{Dickinsonetal12circ}), this object
remains somewhat enigmatic.

Using the C abundance derived from the C\,{\sc iv} doublet at
WD\,1942+499 and WD\,2257$-$073, \cite{Lallementetal11}  predicted a strong C\,{\sc iii} multiplet that was not seen in the \textit{FUSE} spectra of the stars. This was used to infer a non-photospheric
origin for the  C\,{\sc iv} seen at these objects. Coupled with the
contaminating effects detailed here, one can see that the use of the
non-resonance C\,{\sc iii} absorption lines ($1s^22s2p \rightarrow
1s^22p^2$) gives a better indication of photospheric C abundance
than the C\,{\sc iv} resonance transitions ($1s^22p \rightarrow
1s^22s$) where non-photospheric material is present, since the
resonance lines will also be present in any highly ionised, low
density material along the sight line to a given star. Furthermore,
such significant differences in C abundances derived using the C\,{\sc iii} and
C\,{\sc iv} absorption features can be a useful diagnostic of the
presence of circumstellar material in future studies. Indeed, were the
circumstellar and photospheric components to be completely unresolved,
giving rise to symmetric blended absorption features, comparisons of
abundances derived using resonant and non-resonant transitions could
be used to infer the presence of the circumstellar material.

\begin{figure}
\includegraphics[width=0.47\textwidth]{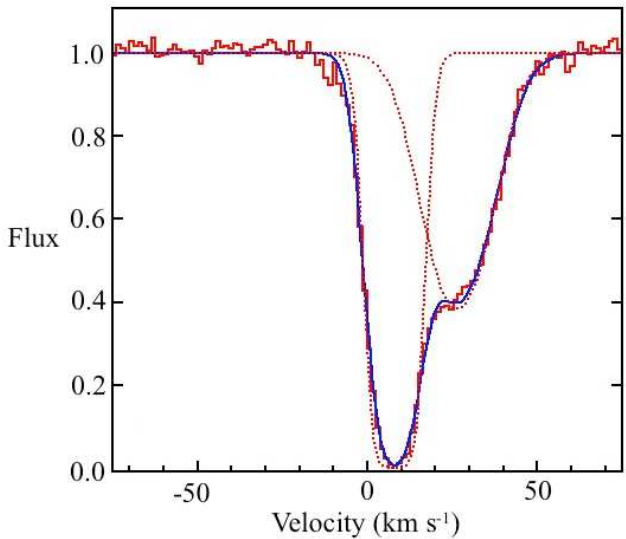}
 \caption{The C {\sc iv} 1548 \AA\ line fit using the method outlined in
   \cite{Dickinsonetal12circ}. The histogram plot represents the observed
   data (red in online copy), while the smooth (blue online) line
   represents the best fitting model. Model components are plotted with
   dotted lines. The heavily absorbed component at 8 km s$^{-1}$ is
   circumstellar, while the photospheric component is found at 26.7 km s$^{-1}$.}
  \label{fig:barry}
\end{figure}

One way to better constrain the abundance of these stars
would be to physically model both the photospheric and circumstellar
line profiles, rather than modelling one and approximating the other,
as has been done here and in previous work. Indeed, the approximation
of the photospheric components used by \cite{Dickinsonetal12circ}
yielded line profiles inconsistent with those predicted
from stellar atmosphere models (an example of this can be seen when
comparing the 1548 \AA\ C\,{\sc iv} line profile of WD\,0501+527 in figure \ref{fig:BFfit}
to that obtained using the fitting method used in \citealt{Dickinsonetal12circ}; figure
\ref{fig:barry}). This will allow a better understanding of both the
photospheric abundances of the hot white dwarfs and the conditions present in the ionised circumstellar medium, and will be the subject of future work. Indeed, a robust physical model of the these absorption features may go some way to better understanding the extremely narrow, almost saturated absorption features of the very hot (110\,000\,K; \citealt{Barstowetal03}) DA WD\,0948+534 (PG\,0948+534), that as yet have proven difficult to model (\citealt{Barstowetal03,Dickinsonetal12nv}).

Of the 16 stars surveyed by
\cite{Bannister03} and \cite{Dickinsonetal12circ} with \textit{T}$\rm _{eff}$
values greater than 50\,000\,K, seven display unambiguous circumstellar
absorption in both studies, with a further four stars (WD\,0621$-$376,
WD\,0948+534, WD\,2211$-$495 and WD\,2331$-$475) showing evidence for possible,
unresolved circumstellar material in one or both studies. This means between 44\,\% and 69\,\% of the hot DAs surveyed have circumstellar lines in their
spectra; re-observation of the 
objects with possible circumstellar material in the Bannister/Dickinson
sample at higher resolution and signal to noise will allow this
fraction to be better constrained. Additionally, of the stars with \textit{T}$\rm _{eff}$
values less than 50\,000\,K, one (WD\,1611$-$084)  displays circumstellar
lines, with two other stars serving as candidates for such material (WD\,0050$-$335 and WD\,2152$-$548). Recently, high signal to noise (S/N)
observations have allowed circumstellar absorbing components
to be detected at other DAs
(e.g. \citealt{Lallementetal11}). Observation of a wider hot DA sample will allow a larger, more
statistically robust sample to be built up. This will more accurately
establish the ubiquity of this
phenomenon, allowing a deeper understanding of
how these circumstellar features affect our picture of the hot white
dwarf photosphere, and
how these stars interact with their circumstellar environments, the ISM
and any material still left from the stars' PN phase, giving crucial insights
into how young, hot white dwarfs and their environments evolve.

\section*{Acknowledgements}
\label{acknowledgements}
N.J.D. and M.A.B. acknowledge the support of STFC. B.Y.W. would like to
acknowledge Guaranteed Time Observer funding for this research through NASA
Goddard Space Flight Center grant 005118. N.J.D. wishes to thank Jay Holberg
and Ivan Hubeny for useful discussions. We thank the referee for useful comments.


\end{document}